\documentclass{IAU-TrA}
\usepackage{graphicx}
\usepackage{natbib}

\title[WORKING GROUP ON MOLECULAR DATA] 
{}

\author[DIVISION~B/COMMISSION~14/WORKING GROUP 
ON MOLECULAR DATA] 
{}

\pubyear{2014}
\volume{Volume XXIXA}
\pagerange{001--008}
\jname{Transactions IAU, Volume XXIXA}
\editors{Thierry Montmerle, ed.}
\begin{document}

\maketitle

{\bf

\large
\noindent
DIVISION B/COMMISSION~14/WORKING GROUP ON                         \\ 
MOLECULAR DATA                                                     \\

\normalsize

\begin{tabbing}
\hspace*{45mm}  \=                                                   \kill
CHAIR           \> Steven R. Federman                                  \\
VICE-CHAIR      \> Peter F. Bernath                                \\
VICE-CHAIR      \> Holger S.P. M\"{u}ller                          \\
\end{tabbing}

\vspace{3mm}

\noindent
TRIENNIAL REPORT 2012-2014
}

\firstsection 

\section{Introduction}

The current report covers the period from the second half of 2011 to late 
2014. It is divided into three areas covering rotational, vibrational, and 
electronic spectroscopy. A signifcant amount of experimental and theoretical work 
has been accomplished over the past three years, leading to the development and 
expansion of a number of databases whose links are provided below. Two notable 
publications have appeared recently: An issue of The Journal of Physical Chemistry 
A in 2013 honoring the many contributions of Takeshi Oka (J. Phys. Chem. A, 117, pp. 
9305-10143); and IAU Symposium 297 on Diffuse Interstellar Bands \citep{cam14}. A 
number of the relevant papers from these volumes are cited in what follows. 
Related research on collisions, reactions on grain surfaces, and astrochemistry 
are not included here. 

\section{Rotational Spectra}
\label{Rotational Spectra}

A large number of reports have appeared dealing with rotational spectra of 
molecules potentially relevant to radio-astronomical observations. Therefore, 
emphasis will be put on investigations dealing with molecules already 
observed in space, and on molecules related to these species. 
A few observational papers are mentioned to highlight some recent findings. 
The grouping of the molecules has been modified with respect to the previous report. 
Complex organic molecules are very important for ALMA, and they have attracted 
considerable attention in the last three years. They form the first group 
and include most of the so-called weed molecules. Molecules which (may) occur 
in circumstellar envelopes of late type stars form the second group. 
Additional groups deal with hydride molecules, with cations, or with other 
molecules. For completeness, we also include determinations of transition 
frequencies from radio astronomical observations, provided these are of 
sufficient importance. 
\smallskip

Several databases provide rotational spectra of (mostly) molecular species 
of astrophysical and astrochemical relevance. The two most important sources 
for predictions generated from experimental data by employing appropriate 
Hamiltonian models are the Cologne Database for Molecular Spectroscopy, 
CDMS\footnote{http://www.astro.uni-koeln.de/cdms/} 
\citep{CDMS_1,CDMS_2} with its 
catalog\footnote{http://www.astro.uni-koeln.de/cdms/catalog} 
and the JPL catalog\footnote{http://spec.jpl.nasa.gov/} 
\citep{JPL-catalog_1998}. Both also provide 
primary information, i.e. laboratory data with uncertainties, mostly 
in special archive sections. Additional primary data are available 
in the Toyama Microwave 
Atlas\footnote{http://www.sci.u-toyama.ac.jp/phys/4ken/atlas/}.
A useful resource on the detection of certain molecular transitions in space 
is the NIST Recommended Rest Frequencies for Observed Interstellar Molecular 
Microwave Transitions, which has been updated some years ago \citep{NIST-RRF_2004}.
\smallskip

The European FP7 project Virtual Atomic and Molecular Data Centre, 
VAMDC\footnote{http://www.vamdc.org/}, 
aims at combining several spectroscopic, collisional, 
and kinetic databases. The CDMS is the rotational spectroscopy database taking 
part\footnote{http://cdms.ph1.uni-koeln.de/cdms/portal/}; 
recently, the JPL catalog 
has joined; several infrared databases are also involved. The project has 
been described by \citet{VAMDC_2010}. 
\smallskip

Other tertiary sources combining data from various databases are, e.g., 
Cassis\footnote{http://cassis.cesr.fr/}, 
which provides tools to analyze astronomical spectra, 
lamda\footnote{http://www.strw.leidenuniv.nl/$\sim$moldata/} 
which also contains collisional data, or 
splatalogue\footnote{http://www.splatalogue.net/}. These databases 
rely heavily on the CDMS and JPL catalogs for their rotational data. This applies 
in part also to the infrared databases HITRAN and GEISA for selected rotational 
or rovibrational data. 

\subsection{Complex molecules}
\label{complex}

An overview of molecular complexity achievable with single dish radio telescopes 
is provided by \citet{SgrB2_survey_2013}. They carried out molecular line surveys
of the giant molecular clouds Sagittarius~B2(N) and (M) close to the Galactic 
Center with the IRAM 30\,m telescope at 3\,mm with additional observations at 2 
and 1.3\,mm. The detections include several new molecules, isotopologues, and 
excited vibrational states. Follow-up ALMA Cycle-0 observations resulted in the 
detection of the first branched alkyl compound, \textit{iso}-C$_3$H$_7$CN 
\citep{i-PrCN_det_2014}. 
\smallskip

\citet{E-Cyanomethanimine_rot_det_2013} extended the rotational specroscopy of 
cyanomethanimine and described the detection of the $E$ conformer, which is 
presumably higher in energy than the $Z$ conformer, but has a much larger 
dipole moment. \citet{EtSH_rot_2014} extended the ethyl mercaptan data sets 
considerably and concluded that the molecule would be present in Orion. 
We also emphasize investigations and astronomical detections of dimethyl ether 
with one D \citep{D-DME_rot_det_2013} or with one or two $^{13}$C 
\citep{13C-DME_rot_det_2013}, as well as methyl formate in its doubly 
torsionally excited state \citep{MeFo_vt=2_rot_2013} or several of its isotopologues 
\citep{Me-D-MeFO_rot_det_2013,13C_MeFo_rot_2014,18O-MeFo_rot_det_2012}. 
There was also an account on the rotational spectroscopy of the very high lying 
\textit{anti}-conformer and its tentative detection in space 
\citep{anti_MeFo_rot_tent_det_2012}. 
\smallskip

Other reports include THz frequency \citep{MeOH_around_2.6THz_2011} and intensity 
studies \citep{MeOH_int-600GHz_2014,MeOH_int_250GHz_2014} of CH$_3$OH, 
extensive studies of $^{13}$CH$_3$OH \citep{13MeOH_rot_2014}, 
CH$_2$DOH \citep{CH2DOH_rot_2012,CH2DOH_rot_2014}, CH$_3$SH \citep{MeSH_rot_2012}, 
and ethyl cyanide \citep{EtCN_v21_v12_rot_det_2013,13C-EtCN_rot_2012}, as well as 
vinyl cyanide \citep{VyCN_up-to-v11=2_rot_2012,VyCN_isos_rot_2011,VyCN_rot_2014}, 
including redetermination of their dipole moment components \citep{EtCN_VyCN_dip_2011} 
with a considerable change in $\mu _b$ for vinyl cyanide.
\smallskip

\citet{acetone_analysis_2013} presented a somewhat restricted reanalysis of the 
rotational spectrum of acetone in its ground and its two singly excited torsional 
states employing a new BELGI variant. This program may be very useful for a variety 
of two-top internal rotor problems.
\smallskip

Several promising molecules can be searched for in space because of greatly extended 
laboratory data, such as butanone \citep{butanone_rot_2013}, \textit{n}-butyl cyanide 
\citep{n-BuCN_rot_2012}, 1,2- \citep{1-2-PD_rot_2014} and 1,3-propanediol 
\citep{1-3-PD_rot_2013}, urea \citep{Urea_rot_obs_2014}, and 
2-aminopropanitrile \citep{2-APN_rot_2012}; data on 3-aminopropanitrile are still 
very limited.
\smallskip

In addition, there are now several minor isotopologues for which searches are now 
viable: e.g. several isotopologues of methanimine \citep{H2CNH_isos_rot_2014} 
and cyanamide \citep{cyanamide_isos_rot_2011}, deuterated formamide 
\citep{D-FA_rot_2013}, $^{13}$C-containing ethanol \citep{13C-EtOH_rot_2012}, 
and deuterated \citep{D_glycald_rot_2012} and $^{13}$C-containing glycolaldehyde 
\citep{13C-glycald_rot_2013}. 
\smallskip

Other studies involve H$_2$CNH \citep{H2CNH_rot_2012}, H$_2$CCNH 
\citep{H2CCNH_rot_2014}, \textit{n}-butanal \citep{n-Butanal_FTMW_2012}, acetic 
acid \citep{HAc_rot_2013}, aminoacetonitrile \citep{AAN_rot_2013} and 
methyleneaminoacetonitrile \citep{MAAN_rot_2013}, methylamine \citep{MeNH2_rot_2014}, 
methyl acetate \citep{MeAc_rot_2014}, acetaldehyde \citep{CH3CHO_rot_2014}, and 
deuterated methyl cyanide \citep{D-MeCN_rot_2013}.

\subsection{Circumstellar molecules}
\label{CS-mols}

In recent years, several molecular species were detected first in O-rich 
circumstellar envelopes (CSEs) of late-type stars. One recent prominent example 
is TiO$_2$, a dust-forming molecule, detected in the CSE of VY~CMa \citep{det_TiO2_2013}.
Initially, detections of molecules in CSEs were commonly made toward C-rich stars, 
in particular CW~Leonis, also known as IRC~+10216. Remarkable examples are the 
laboratory spectroscopy and detection of HMgNC \citep{HMgNC_rot_det_2013} and 
the rotational spectroscopy of FeCN and FeNC \citep{FeCN_FeNC_rot_2011}; 
the detection of FeCN was reported earlier, though. 
\smallskip

Investigations of detected species involve $^{29}$SiC$_2$ and $^{30}$SiC$_2$ 
\citep{29_30SiC2_rot_2011} and SiO \citep{SiO_rot_2013} along with an analysis 
of recent NaCN data \citep{NaCN_parameter_2012}. SiC$_2$ spectroscopic parameters 
were improved once more through radio astronomical observations 
\citep{SiC2_astro_rot_2012}.
\smallskip

Further investigations include ScS and YS \citep{ScS_YS_FTMW_2012}, $^{25}$MgH 
and $^{67}$ZnH \citep{25MgH_67_ZnH_rot_2013}, KSH \citep{KSH_rot_2013}, AlC$_3$N 
\citep{AlC3N_rot_2014}, PCN \citep{PCN_rot_2012}, PH$_2$CN \citep{PH2CN_rot_2014}, 
SiC$_3$H \citep{SiC3H_rot_2014}, SiC$_2$N and SiC$_3$N \citep{SiC2_3N_rot_2014}, 
alkali metal acetylides \citep{Alkali-CCH_2011_FTMW_2011}, 
and AlCCH \citep{AlCCH_rot_2012}. 
\citet{AlH_rot_2014} reported a transition frequency for the AlH $J = 2 - 1$ 
transition which is not compatible with numerous IR and optical studies. 
Therefore, we recommend to avoid using this datum until further clarification. 

\subsection{Hydrides}
\label{hydrides}

Hydrides are here all molecules consisting of one non-metal atom and one or more 
H atoms. They may be neutral or charged. Metal hydrides have been dealt with in 
subsection~\ref{CS-mols}. Even though the \textit{Herschel} mission was terminated 
late in April 2013, data are still being analyzed; 
for instance, quite recently the detection 
of ArH$^+$ was reported \citep{ArH+_det_2013,ArH+_diff_2014}. Moreover, the German 
REceiver At Terahertz frequencies (GREAT) on board of the Stratospheric Observatory 
For Infrared Astronomy (SOFIA) has opened new observing capabilities above 1~THz. 
In addition, transitions of heavier hydrides or higher rotationally excited 
transitions of lighter asymmetric top hydrides, such as H$_2$O, can be observed 
from the ground and may be of particular importance for ALMA.
\smallskip

Noteworthy is especially the investigation of HCl$^+$ \citep{HCl+_rot_2012}, 
which was important to establish its detection in the ISM. We point out also 
extensive studies of H$_2$O 
\citep{H2O_up-to-1st-triade_rot_2012,H2O_2nd-triade_rot_2013,H2O_high-J_rot_2014}. 
\citet{HD_H2O_NH3_around_2.6THz_2011} reported not only transition frequencies 
of H$_2$O around 2.6~THz, but also of HD and NH$_3$. Noteworthy are also the 
extensive investigations on sulfur isotopologues of H$_2$S 
\citep{H2S_div-isos_FIR_2013,H2S_rot_2014,H2S-36_rot_2014} and those of H$_2$F$^+$ 
\citep{H2F+_rot_2012}. Although H$_2$F$^+$ is predicted to have small column densities 
in space, it is necessary to know selected transitions, in particular those involving 
the ground states, well enough.
\smallskip

Other studies include investigations of H$_3$O$^+$ \citep{H3O+_rot_2014}, NH$_2$ 
\citep{NH2_FIR_2014}, $^{15}$NH \citep{15NH_rot2012}, several OH isotopologues 
\citep{OH_div-isos_rot_2013}, OH$^-$ \citep{OH-_rot_2014}, CH 
\citep{CH_etc_FIR_2011,CH_MW_2014}, and an analysis of PH$_3$ rotational data 
\citep{PH3_analysis_2013}. \citet{SH+_obs+_parameter_2014} determined improved 
spectroscopic parameters for SH$^+$ based on radio astronomical observations 
of the lowest frequency transitions. 

\subsection{Cations}
\label{cations}

Cationic light hydrides were mentioned in subsection~\ref{hydrides}. With regard 
to heavier cations, the laboratory spectroscopy of C$_3$H$^+$ \citep{C3H+_rot_2014} 
is particularly worth mentioning as it confirmed the earlier assigment of a long 
series of U-lines detected in the Horsehead Nebula mentioned in our last report. 
\citet{H2NCO+_rot_det_2013} reported on the rotational spectrum of H$_2$NCO$^+$ 
and its tentative detection in Sgr~B2(N). 
\smallskip

Other studies involve CO$^+$ \citep{CO+_rot_2013}, HNNO$^+$ \citep{HNNO+_rot_2013}, 
HOSO$^+$ \citep{HOSO+_rot_2011}, C$_2$H$_3$CNH$^+$ \citep{prot-VyCN_rot_2013}, and 
Terahertz spectra of N$_2$H$^+$, HCO$^+$, and CF$^+$ \citep{N2H+_HCO+_CF+_rot_2012}.

\subsection{Other molecules}
\label{other}

We emphasize a study of several isotopologues of $c$-C$_3$H$_2$ 
\citep{c-C3H2_div-isos_rot_2012}, which led to the detection of $c$-C$_3$D$_2$. 
Other investigations, which may be worthwhile mentioning, include HCOOD and DCOOH 
\citep{DCOOH_HCOOD_rot_2011}, the vinoxy radical \citep{H2CCHO_rot_2014}, HOOD 
(however with no tranistion frequencies) \citep{HOOD_FIR_2012}, phenol 
\citep{PhOH_rot_2013}, HOSO \citep{HOSO_rot_2013}, and an analysis of several 
O$_2$ isotopologues in several electronic states \citep{O2-analysis_2012}.

\section{Vibrational Spectra}
\label{Vibrational Spectra}

The vibration-rotation spectra of molecules of astronomical or of potential 
astronomical interest are reviewed for the period 2011-2014 starting from the end 
of our previous report \citep{fed12}. In addition to the references to 
particular molecules given below, there are a number of spectral database 
compilations that are useful. Perhaps the most helpful is the HITRAN database that 
contains vibration-rotation line parameters for a large number of species such as 
H$_2$O, CO$_2$, CO, HF and so forth, found primarily in the Earth's atmosphere. A 
new edition has appeared (HITRAN 2012, Rothman et al. 2013); further information 
including updates and corrections is 
available\footnote{http://www.cfa.harvard.edu/hitran/}. 
The molecular coverage is slowly being expanded to cover planetary atmospheres other 
than Earth and HITRAN 2012 contains line parameters for PH$_3$, H$_2$, and CS. HITRAN 
is widely used for astronomical applications although it is not always suitable 
because of missing lines and bands, particularly in the near infrared region. For 
high temperature applications, the HITEMP database \citep{rot10} for H$_2$O, 
CO$_2$, CO, NO, and OH is more suitable and a new edition is being prepared.
\smallskip

For larger molecules, individual vibration-rotation lines are no longer clearly 
resolved and it becomes necessary to replace line-by-line calculations by absorption 
cross sections. The main drawback to using cross sections is that a considerable 
number of laboratory measurements are needed to match the temperature and pressure 
conditions of the objects under observation. HITRAN also includes a number of high 
resolution infrared absorption cross sections for organic molecules such as methanol, 
ethane, and acetone, but the broadening gas is air rather than H$_2$, N$_2$, or CO$_2$. 
While the GEISA database has significant overlap with HITRAN, it contains additional 
molecules of interest for studies of planetary atmospheres \citep{jac11}. 
\smallskip

There are a number of web sites that have collections of spectroscopic line lists or 
infrared absorption cross sections that are updated regularly. The ExoMol 
site\footnote{http://www.exomol.com/} 
of J. Tennyson has an extensive collection of calculated 
line lists designed ``as input to atmospheric models of exoplanets, brown dwarfs and 
cool stars.'' G. Villanueva's 
site\footnote{http://astrobiology.gsfc.nasa.gov/Villanueva/spec.html} 
provides line lists for the 
simulation of infrared emission spectra of comets for species such as ethane, methanol, 
ammonia, and water excited by solar radiation. A very useful set of infrared absorption 
cross sections for several hundred molecules are available from Pacific Northwest 
National Laboratory\footnote{http://nwir.pnl.gov}, PNNL, for the 600-6500 cm$^{-1}$ 
(1.54-16.7 $\mu$m) range \citep{sha04}. While the PNNL spectra are not always suitable 
for astronomical applications because they are recorded with 1 atm of nitrogen as a 
broadening gas at sample temperatures of 278, 293, and 323 K, they can be very useful.
\smallskip

Other interesting sources of infrared data are the ACE high resolution spectral 
atlases of the Sun \citep{has10} and of the Earth's atmosphere \citep{hug14} recorded 
using a high resolution Fourier transform spectrometer in low Earth orbit. The ACE 
atmospheric spectra are recorded by solar occultation (i.e., using the Sun as a light 
source during sunrise and sunset). This geometry matches that used for exoplanet 
transit spectroscopy \citep{ber14} so these spectra are a template for infrared 
absorption spectra of Earth-like planets.
 
\subsection{Diatomic molecules}
\label{diatomics}

Hydrogen is the most abundant element in the Universe so it is no surprise that small 
diatomic hydrides are also abundant. One of the major successes of the 
\textit{Herschel Space Observatory} was the detection of 
many diatomic hydrides, often for the first time, by 
rotational spectroscopy \citep{ben13}. Diatomic hydrides have large rotational constants 
so their pure rotational spectra fall in the Terahertz region where 
absorption by the Earth's atmosphere is a problem. 
With \textit{Hershel} no longer operational, 
infrared observations from the ground with instruments such as Phoenix, CRIRES, and 
TEXES are attractive. The new EXES instrument on the SOFIA (Stratospheric Observatory 
for Infrared Astronomy) aircraft is particularly promising because it flies in the 
stratosphere above most of the Earth's atmosphere.
\smallskip

The argonium ion ($^{36}$ArH$^+$) was detected in the Crab Nebula by \citet{ArH+_det_2013} 
using \textit{Herschel}. On Earth the major isotope is $^{40}$Ar from 
radioactive decay of $^{40}$K in rocks while $^{36}$Ar is the main isotope in stars 
because of nucleosynthetic production in supernova explosions. Improved 
vibration-rotation spectra of the 1-0 band of $^{36}$ArH$^+$ and $^{38}$ArH$^+$ were 
reported by \citet{cue14}. Improved line positions have also been measured for 
HeH$^+$ by \citet{per14}.
\smallskip

The vibration-rotation line parameters of the hydrogen halide molecules (HF, HCl, HBr, 
and HI) have been revised in HITRAN 2012. Of particular interest in astronomy are the 
extensive lists of new line positions and intensities for HF and HCl that are used to 
obtain fluorine \citep{jon14} and chlorine abundances. A similar extensive revision of 
the CH, NH, and OH line parameters has been carried out. These new line 
lists were created using reliable dipole moment functions and inclusion of the 
Herman-Wallis effect using LeRoy's LEVEL computer 
program\footnote{http://leroy.uwaterloo.ca/programs/}. The CH line list of \citet{mas14}
primarily focuses on electronic transitions but also includes vibration-rotation 
bands. The NH \citep{bro14a} and OH \citep{bro15} work adds line 
intensities to our previous analyses of line positions based partly on the ACE solar spectrum.
\smallskip

The ExoMol line lists include BeH, MgH, CaH \citep{yad12}, SiO \citep{bart13}, 
NaCl, KCl \citep{bart14}, and PN \citep{yor14}. These new line lists combine experimental 
measurements and $ab~initio$ calculations. Salt vapors such as NaCl 
and KCl are predicted to be present in hot super-Earth exoplanets \citep{sch12}.  
The infrared bands of SiO are readily observed in K-M giant and supergiant 
stars \citep{ohn14}.
\smallskip

The vibration-rotation bands of H$_2$ are forbidden by electric dipole selection 
rules, but are observable by weak electric quadrupole transitions. The line 
intensities and line positions of the vibration-rotation bands of H$_2$ were 
recalculated for HITRAN 2012 based on $ab~initio$ results \citep{rot13}. The 
line positions are estimated to have an accuracy of about 0.001 cm$^{-1}$ and more 
recent results from calculations \citep{pac14} and experiments \citep{che12}
have even higher accuracy. The HD molecule is polar and so has 
dipole-allowed transitions, which are also reported in HITRAN 2012.
\smallskip

The CN radical is found in a very wide range of sources mainly by radio and 
optical/IR astronomy. However, vibration-rotation lines can also be detected 
\citep{wie91} and new line lists for CN that include the vibration-rotation 
bands for CN \citep{bro14b}, $^{13}$CN, and C$^{15}$N \citep{sne14} have 
been generated with an $ab~initio$ dipole moment function calculated by D. Schwenke 
(NASA-Ames). These extensive line lists are based on recent laboratory observations 
of the $B^2\Sigma^+ - X^2\Sigma^+$ and $A^2\Pi - X^2\Sigma^+$ Violet System and Red 
System, which extends into the near IR spectral region. A new line list has been 
completed for the corresponding infrared $A^2\Pi - X^2\Sigma^+$ electronic transition 
of the isovalent CP radical \citep{ram14a}. The Ballik-Ramsay and Phillips Systems 
of C$_2$ are also prominent in the near infrared and a new perturbation analysis has 
demonstrated that the singlet-triplet splitting was in error by 3 cm$^{-1}$ 
\citep{chen15}. This analysis has astronomical implications for example in the 
excitation of C$_2$ in comets by solar radiation. 
\smallskip

An important application of molecular line parameters is to extract elemental 
abundances from the near infrared spectra of large numbers of cool stars observed in 
surveys such as APOGEE (Apache Point Observatory Galactic Evolution Experiment). 
APOGEE is recording H-band (1.51-1.69 $\mu$m) spectra of thousands of evolved, 
late-type stars with a focus on red giants with a surface temperature of 3400-5000 K 
\citep{cot14}. CNO abundances are derived from CO and OH overtone spectra, 
and the Red System of CN. The improved line parameters for OH and CN, along with 
HITEMP values for CO cited above, are recommended for abundance analyses. Stellar models 
also require dissociation energies and significant improvements have been made using 
the method of Active Thermochemical Tables\footnote{http://atct.anl.gov/}.  For example, 
the latest values for the dissociation energies $D_0$ for C$_2$, CH, CO, CN, and OH are 
(in eV) 6.24475, 3.57154, 11.11092, 7.72400, and 4.41129, respectively \citep{rus14}. 

\subsection{Small polyatomic molecules}
\label{polyatomics}

The line parameters of ammonia \citep{dow13} and methane as given in the HITRAN 
2012 database are satisfactory for most astronomical purposes at low temperatures, 
except for overtone and combination bands in the near infrared and visible regions. 
In the near infrared, \citet{sun12} have generated an empirical line list for 
NH$_3$ covering the 6300 to 7000 cm$^{-1}$ region. By using spectra recorded over a 
range of sample temperatures (185-296 K), empirical lower state energies were 
obtained, although most of the lines still lack detailed quantum number assignments. 
Similar work in Grenoble \citep{cam13} on CH$_4$ has provided the WKLMC 
empirical line lists (5852-7919 cm$^{-1}$) using two temperatures (80 K and 296 K). 
The WKLMC methane line list is a major improvement on the band models typically used 
by planetary astronomers. New measurements for methane spectra in the 4800-5300 
cm$^{-1}$ region have been reported by \citet{nik14}. For PH$_3$ (long 
detected in Jupiter and Saturn) a new analysis for the 5 bands that comprise the 
pentad region between 1950 and 2450 cm$^{-1}$ has appeared \citep{dev14}. Work 
on high overtone and combination bands of CO$_2$ continues \citep{pet13, lu13} 
for applications for Venus and Mars.
\smallskip

The spectra of hot molecules needed to simulate the spectra of cool stars, brown 
dwarfs, and exoplanets remains a challenge. For hot water, calculations continue to 
improve \citep{pol13}, but the BT2 line list \citep{bar06} remains 
the standard for astronomical applications. A new extensive compilation of the 
vibration-rotation energy levels of water has appeared \citep{ten13}. For hot 
ammonia, laboratory spectra recorded in emission are available \citep{har11, har12a} 
and with at least two groups providing rather good calculated spectra 
\citep{hua11a, hua11b, yur11}. There is much recent progress on 
the spectroscopy of hot methane with the experimental line lists of \citet{har12b} 
and two comprehensive calculated line lists \citep{yur14, rey14a};  
a calculation for CH$_3$D has also been carried out \citep{rey14b}. 
For CO$_2$, HITRAN 2012 for cold molecules and HITEMP, CDSD-4000 \citep{tas11} 
and calculations for 13 isotopologues for hot molecules \citep{hua14} are 
recommended. New internal partition functions for NH$_3$ and PH$_3$ \citep{sou14} 
have been published. A new line list for HCN and HNC has been prepared 
by \citet{bar14} combining both experimental and theoretical work.  
\smallskip

The pure rotational spectrum of NH$_3$D$^+$ has been tentatively identified in Orion 
\citep{cer13}, based on a new infrared laboratory spectrum \citep{dom13}. 
H$_2$Cl$^+$ has been detected by \textit{Herschel}, but not H$_2$F$^+$; new 
infrared spectra of H$_2$F$^+$ have been measured with a Fourier transform 
spectrometer \citep{fuj13}.
\smallskip

The carbon chain molecules C$_3$ and C$_5$ can be detected in the circumstellar 
envelopes of carbon stars by infrared observations \citep{har14}. Improved laboratory 
spectra of C$_3$ have been reported in the 3 $\mu$m region \citep{kri13} 
and high resolution photoelectron spectroscopy has been used to refine the 
vibrational frequencies of C$_5$ \citep{wei13}.

\subsection{Large molecules}
\label{large}

There is continuing strong interest in large carbon-containing molecules such as 
C$_{60}$ (e.g., Bern\'{e} \& Tielens 2012) and polycyclic aromatic hydrocarbons 
(PAHs). For PAH molecules the NASA-Ames database has been updated and new features 
added \citep{boe14}. Extensive calculations by the NASA-Ames group have 
continued, for example, on dehydrogenated PAHs \citep{mac14} and PAH clusters 
\citep{ric13}. High resolution infrared absorption spectra of the $\nu_{68}$ 
mode of pyrene near 1184 cm$^{-1}$ were recorded by \citet{bru12}. ZEKE photoelectron 
spectroscopy was used to measure the frequencies of the benzoperylene 
cation \citep{zha12}. There is continuing discussion on the nature of these 
`PAH' bands in astronomical sources, for example with the suggestion that mixed 
aromatic/aliphatic organic nanoparticles, MAONs, rather than free flying PAHs are the 
carrier \citep{kwo13}. Improved infrared absorption spectra of C$_{60}^+$ and 
C$_{60}^-$ in neon matrices have been measured by \citet{ker13} and C$_{60}^+$ 
has been identified in emission in the interstellar medium \citep{bern13}.

\section{Electronic Spectra}
\label{Electronic Spectra}

Recent work on electronic spectra, such as line identification, energy levels, and 
related data needed for photochemical models, are described. These data include 
absorption cross sections (or equivalently lifetimes, transition probabilities, and oscillator strengths), predissociation widths and rates, and analyses of 
anomalies in line strength and width caused by perturbations between energy levels. 
Both empirical (experimental and astronomical) and theoretical results are 
presented. The section is divided into four topics: interstellar matter, including 
diffuse molecular clouds, disks around newly formed stars, and comets whose chemistry 
is similar; metal hydrides and oxides in the atmospheres of late-type stars; the 
atmospheres of planets and their satellites; and larger molecules. Although some of 
the work is noted in the sections on rotational and vibrational spectra, electronic 
spectroscopy is stressed here.

\subsection{Interstellar matter}
\label{Interstellar matter}

Because observations and analyses of CO and its photochemistry is central to 
astrophysical studies, a large body of new work has appeared since the last report. 
Oscillator strengths and predissociation rates with improved precision are now 
available for a number of transitions in CO isotopologues 
\citep{eid12, eid14, hea14a, sta14}; another study focused on self shielding among 
the isotopologues \citep{cha12}. Since CO photodissociation involves line absorption, 
self shielding, which arises when dissociating transitions become optically thick, 
allows molecules in the cloud interior to be protected. A theoretical study on 
predissociation in the $E$ state \citep{maj14a} obtained a line width consistent 
with earlier measurements. Using a newly developed technique, Ng and colleagues 
\citep{gao11a, gao12, gao13a, gao13b} measured branching fractions for the atomic 
products arising from dissociation for specific rotational levels. Branching 
fractions were also obtained for other systems, including CO$_2$ and N$_2$ 
\citep{gao11b, pan11, lu14} as well as O$_2$ \citep{hol12, zho14}. Similar studies 
involve N$_2$ photoionization \citep{hol12, oke12} and CO$_2$ photoionization 
\citep{fur13}.
\smallskip

New measurements of the \r{A}ngstr\"{o}m ($B^1\Sigma^+ - A^1\Pi$) and fourth 
positive ($A^1\Pi - X^1\Sigma^+$) systems improved our knowledge of the 
perturbations affecting the $A$ state. \citet{hak12} and \citet{hak13} studied 
\r{A}ngstr\"{o}m system bands in the rare $^{13}$C$^{17}$O isotopologue; Hakalla 
and colleagues obtained spectra for these bands in other isotopologues 
\citep{hak12a, hak12b, kep14, hak14}. As for the $A$--$X$ system of bands, 
\citet{kep11} and \citet{gav13} provided further details on the perturbations in 
$^{13}$C$^{16}$O, while \citet{niu13} analyzed $^{12}$C$^{16}$O spectra. Lifetimes 
for the main isotopologue were obtained by \citet{blo11}. A time-dependent quantum 
mechanical study \citep{maj12} yielded oscillator strengths; the slight differences 
with experimental results likely arise from perturbations in the $A$ state. 
Moreover, accurate line positions were measured for transitions that are sensitive 
to the proton-to-electron mass ratio \citep{nij11, sal12, niu15}, and \citet{nij13} 
tested mass-scaling relations among isotopologues for the $a^3\Pi$ state.
\smallskip 

Since N$_2$ and CO are isoelectronic, photochemical models 
of interstellar environments need to take this into account \citep{hea14b}. Much 
like the studies on CO discussed above, new experimental results on oscillator 
strengths and predissociation \citep{hea11, wu12} were reported. Cross sections for 
electron-impact excitation \citep{mal12} and photoionization efficiencies for N$_2$ 
isotopologues \citep{ran14} were obtained as well. \citet{lit13} studied the singlet 
\and triplet states of N$_2$ through $ab~initio$ calculations. \citet{wu13} 
characterized far ultraviolet absorption of N$_3$ and N$_2^+$ in an N$_2$ matrix.

Other simple carbon-bearing molecules received attention in the past three years. 
Measurements on the Swan band system ($d^3\Pi - a^3\Pi$) in C$_2$ produced 
new molecular constants for several bands \citep{chan13, yeu13, bor13} and line 
strengths for others including the $^{12}$C$^{13}$C isotopologue \citep{bro13, ram14b}. 
Higher-lying vibrational levels in the $X$, $A$, $a$, and $d$ states were observed by 
\citet{nak13}; they also analyzed perturbations seen in the $v$ $=$ 8 level of the 
$d$ state. \citet{nak14} described further studies of the $d^3\Pi_g - c^3\Sigma_u^+$ 
band system. \citet{hup12} analyzed perturbations involving the $F$ 
state that were seen in astronomical spectra. An $ab~inito$ study by \citet{sch11} 
provided spectroscopic constants for a newly identified quintet state. The 
$A^2\Pi - X^2\Sigma^+$ and $B^2\Pi - X^2\Sigma^+$ band systems in CN 
and its isotopologues were studied by Bernath and colleagues. Spectroscopic results 
on $^{12}$C$^{15}$N \citep{col12}, $^{13}$C$^{14}$N \citep{ram12}, and $^{13}$C$^{15}$N 
\citep{col14} were published, and line strengths in the form of oscillator strengths 
were derived for $^{12}$C$^{14}$N \citep{bro14b} and its isotopologues \citep{sne14}. 
Hyperfine structure in spectra of the $A$--$X$ band was discussed by \citet{for14}. 
Recent theoretical efforts on CN involved molecular properties of its low-lying 
electronic states \citep{shi11a} and the determination of photodissociation cross 
sections \citep{elq13}. Two studies on C$_3$ appeared; one provides laboratory 
spectra on $A$--$X$ bands for C$_3$ and its isotopologues \citep{had14} and the 
other gives oscillator strengths for a number of these bands based on astronomical 
measurements \citep{sch14}.
\smallskip

Studies also focused on other species found in interstellar clouds and comets. 
Theoretical work appeared on photoabsorption and photodissociation for H$_2$ 
\citep{mez14}, HeH$^+$ \citep{lor13, miy11}, ArH$^+$ \citep{rou14}, LiH$^+$ 
\citep{bov11}, NH$_3$ \citep{cha13}, and H$_2$O \citep{jia12, zhou14}. Low-lying 
states in CS were studied theoretically by \citet{shi11b}, and \citet{pon14} 
reported theoretical calculations on CS photoionization. Calculations on low-lying 
states of HCl \citep{eng12} were performed as well. Experimental lifetimes on the 
\textit{\~{A}}$^2A_1$ state of NH$_2$ were measured by \citet{ndo12}. Photoionization 
cross sections for CO, N$_2$, and H$_2$O were also determined theoretically 
\citep{rub14}.  It is also worth noting a compilation of transition probabilities for 
several diatomic species \citep{bil14}.

\subsection{Late-type stars}
\label{Late-type stars}

During this reporting period, new results on MgH, FeH, and ZrO were published. 
\citet{hin13} identified lines from isotopologues of MgH associated with the 
$A$--$X$ system, and \citet{gha13} provided transition probablities for lines of 
the $A$--$X$ and $B^{\prime}$--$X$ systems for the main isotopologue. Furthermore, 
\citet{zha14} conducted Zeeman spectroscopy on the $A$--$X$ (0,0) band. An 
$ab~initio$ study of the $A$ and $B^{\prime}$ states of MgH also yielded transition 
probabilities \citep{mos12}. Low-lying electronic states of FeH were described by 
theory \citep{dey12}, as were the $e^1\Pi - X^1\Sigma^+$ and 
$^1\Sigma^+ - X^1\Sigma^+$ systems of ZrO \citep{sha11}, where the latter study 
presented data including oscillator strengths.

\subsection{Planetary atmospheres}
\label{Planetary atmospheres}

The molecules CO$_2$ and SO$_2$ continue to draw attention by experimentalists and 
theorists. Photoabsorption cross sections for electronic transitions in CO$_2$ were 
obtained through experimental measurements \citep{arc13, ven13} and theoretical 
calculations \citep{gre12, gre13}. Low-lying electronic states were also studied 
theoretically \citep{zho13}. As for SO$_2$, excited state dynamics were studied 
experimentally \citep{wil14} and theoretically \citep{mai14, lev14}, while \citet{xie13} 
determined potential energy surfaces for the two lowest singlet and two lowest triplet 
states. 
\smallskip

Data needs for other molecules were also addessed. Absorption cross sections for 
isotopologues of SO were obtained theoretically by \citet{dan14}. A global fit 
to transitions across the spectrum for O$_2$ yielded isotopically invariant data 
for the $X$, $a$, and $b$ states \citep{O2-analysis_2012}, from which an detailed 
analysis of the airglow bands was accomplished \citep{dro13}. Measurements on 
absorption cross sections for C$_2$H$_2$ \citep{che11} were also reported.

\subsection{Larger Molecules}
\label{Larger Molecules}

Most studies of electronic transitions in large molecules seek correspondences 
with wavelengths associated with diffuse interstellar bands. Here we give a sampling 
of the efforts in this area. PAH-related species have drawn considerable attention. 
Recent spectroscopic measurements included results on neutral species by \citet{gre11} 
and on ionized species by \citet{gar11}, \citet{bon11}, and \citet{har13}. A combined 
experimental and theoretical study on propadienylidene (C$_3$H$_2$) was conducted by 
\citet{sta12}. \citet{chak13} measured the spectrum of the triacetylene cation. A 
theoretical computation by \citet{maj14b} provided the spectrum of 
the cyanomethyl anion (CH$_2$CN$^-$). The photoionization of cyanoacetylene was also 
the focus of an experimental study \citep{lea14}.

\vspace{3mm}
 
{\hfill Steven R. Federman}

{\hfill {\it chair of Working Group}}

\end{document}